\documentclass[9pt, conference]{IEEEtran}
\IEEEoverridecommandlockouts
\usepackage{multirow}
\usepackage{cite}
\usepackage{amsmath,amssymb,amsfonts}
\usepackage{algorithmic}
\usepackage{graphicx}
\usepackage{textcomp}
\usepackage{xcolor}
\usepackage{hyperref}
\usepackage{subfigure}

\usepackage{url}
\usepackage{booktabs, tabularray}
\usepackage{tabularx}

\def\BibTeX{{\rm B\kern-.05em{\sc i\kern-.025em b}\kern-.08em
    T\kern-.1667em\lower.7ex\hbox{E}\kern-.125emX}}
\begin{document}

\title{SoloAudio: Target Sound Extraction with Language-oriented Audio Diffusion Transformer}


\author{\IEEEauthorblockN{Helin Wang\IEEEauthorrefmark{2}\thanks{\IEEEauthorrefmark{2} Indicates equal contribution.},
Jiarui Hai\IEEEauthorrefmark{2},
Yen-Ju Lu,
Karan Thakkar,
Mounya Elhilali, and
Najim Dehak
}
\IEEEauthorblockA{Department of Electrical and Computer Engineering, Johns Hopkins University, Baltimore, MD, USA}
\IEEEauthorblockA{Email: hwang258@jhu.edu, jhai2@jhu.edu}
}

\maketitle

\begin{abstract}
In this paper, we introduce SoloAudio, a novel diffusion-based generative model for target sound extraction (TSE). Our approach trains latent diffusion models on audio, replacing the previous U-Net backbone with a skip-connected Transformer that operates on latent features.
SoloAudio supports both audio-oriented and language-oriented TSE by utilizing a CLAP model as the feature extractor for target sounds. Furthermore, SoloAudio leverages synthetic audio generated by state-of-the-art text-to-audio models for training, demonstrating strong generalization to out-of-domain data and unseen sound events.
We evaluate this approach on the FSD Kaggle 2018 mixture dataset and real data from AudioSet, where SoloAudio achieves the state-of-the-art results on both in-domain  and out-of-domain data, and exhibits impressive zero-shot and few-shot capabilities.
Source code\footnote{\href{https://github.com/WangHelin1997/SoloAudio}{https://github.com/WangHelin1997/SoloAudio}} and demos\footnote{\href{https://wanghelin1997.github.io/SoloAudio-Demo}{https://wanghelin1997.github.io/SoloAudio-Demo}} are released.
\end{abstract}


\section{Introduction}

Human beings possess the remarkable ability to focus on a specific sound within a complex acoustic scene composed of various overlapping sound events \cite{DBLP:journals/taslp/DelcroixVOKOA23,DBLP:conf/icassp/ChongWZZ23}. 
Recent works that aim to replicate this human capability computationally have framed the task as target sound extraction (TSE) \cite{DBLP:journals/taslp/DelcroixVOKOA23,DBLP:conf/interspeech/WangYWYZ22,DBLP:conf/interspeech/DelcroixVOKA21,DBLP:conf/icassp/KimBKC24}. 
The objective of TSE is to extract sounds of interest from mixtures of overlapping audio, guided by clues that provide information about the target sound class. These clues can take the form of one-hot labels \cite{DBLP:conf/interspeech/OchiaiDKIKA20,DBLP:conf/interspeech/WangYWYZ22}, audio clips \cite{DBLP:conf/icassp/GfellerRT21}, or images \cite{DBLP:conf/icassp/LiQCWYLQZ23,DBLP:conf/iccv/GaoG19}.

Most prior methods are based on discriminative models, 
which aim to minimize the difference between the estimated and target audio \cite{DBLP:conf/icassp/VeluriCICYG23,DBLP:conf/interspeech/DelcroixVOKA21}. 
While these models often produce good separation in non-overlapping regions, they tend to suffer significant performance degradation in overlapping areas. This is especially problematic in real-world scenarios where sound overlaps are common, making it a critical issue to address in TSE.
With the advent of denoising diffusion probabilistic models (DDPMs) \cite{ho2020denoising,DBLP:conf/iclr/ZhangTC23}, generative models have recently been applied successfully to TSE and source separation tasks \cite{wang2024noise,DBLP:conf/icassp/HaiWYTDE24,DBLP:conf/iclr/MarianiTPMCR24,DBLP:conf/interspeech/KamoDN23}. DPM-TSE \cite{DBLP:conf/icassp/HaiWYTDE24}, a generative approach based on DDPM, achieves both cleaner target renderings and improved separability from unwanted sounds compared to discriminative models.
However, DPM-TSE operates on log-mel spectrograms, where the diffusion process is applied, inherently limiting the reconstruction quality. Additionally, DPM-TSE relies solely on in-domain one-hot labels, which restricts its ability to generalize to out-of-domain data and unseen sound events.

Another challenge in the TSE task is the scarcity of training data, particularly clean, single-label audio, which is often used as the ground truth for target sounds. AudioSep \cite{DBLP:journals/corr/abs-2308-05037} trains open-domain audio source separation models using natural language queries, with mixtures of large-scale multi-label audios. However, it struggles to produce clean sound isolations for a single target sound, which is critical for real-world applications.

To address these issues, we propose SoloAudio, an audio- and/or language-oriented diffusion Transformer model for TSE. Our main contributions are summarized as follows:\\
(i) We introduce a novel Transformer backbone with skip connections, applying the diffusion process in the latent space of an audio variational autoencoder (VAE). SoloAudio supports both audio clues and text clues, by utilizing a CLAP model \cite{DBLP:conf/icassp/wu2023large}.\\
(ii) We leverage synthetic audio from a text-to-audio (T2A) generation model \cite{DBLP:conf/icassp/kong2024improving} as additional training data. Thanks to advancements in T2A, we can generate high-quality, clean audio to improve the training of TSE models.\\
(iii) Experimental results on mixtures from the FSD Kaggle 2018 dataset \cite{DBLP:conf/ismir/FonsecaPFFBFOPS17}  demonstrate that SoloAudio significantly outperforms state-of-the-art methods. Moreover, SoloAudio exhibits strong zero-shot and few-shot capabilities on out-of-domain data and unseen sound events.\\
(iv) Subjective evaluations on real-world data consistently demonstrate a clear preference among listeners for the audio extracted by SoloAudio, highlighting its superior ability to isolate target sounds while effectively eliminating irrelevant noise.

\section{Methodology} 
\subsection{Denoising Diffusion Probabilistic Model (DDPM)}
DDPMs consist of a forward and backward process. The forward process incrementally adds Gaussian noise to the data, following a variance schedule $\beta_1, \ldots, \beta_T$.
\begin{equation}
q\left(x_t \mid x_{t-1}\right):=\mathcal{N}\left(x_t ; \sqrt{1-\beta_t} x_{t-1}, \beta_t \mathbf{I}\right)
\end{equation}

The forward process enables sampling $x_t$ at any
timestep $t$ in a closed form ($x_0$ is the clean signal):
\begin{equation}
x_t=\sqrt{\bar{\alpha}_t} x_0+\sqrt{1-\bar{\alpha}_t} \epsilon,
\end{equation}
where $\alpha_t=1-\beta_t$, $\bar{\alpha}_t:=\prod_{s=1}^t \alpha_s$, and $\epsilon \sim \mathcal{N}(\mathbf{0}, \mathbf{I})$.

Following \cite{DBLP:conf/icassp/HaiWYTDE24},
we use a modified diffusion scheduler and $v$ prediction to improve the purity and overall performance of sound extraction. Additionally, we implement a diffusion noise schedule by keeping \(\sqrt{\bar{\alpha}_{1}}\) unchanged, changing \(\sqrt{\bar{\alpha}_{T}}\) to zero, and linearly rescaling \(\sqrt{\bar{\alpha}_{t}}\) for intermediate \(t \in[2, \ldots, T-1]\) respectively. This adjustment resolves the mismatch between training and inference and prevents the introduction of additional noise during sampling.
A neural network is applied to predict velocity $v_t$:
\begin{equation}
v_t =\sqrt{\bar{\alpha}_t} \epsilon-\sqrt{1-\bar{\alpha}_t} x_0,
\end{equation}


In the reverse process of diffusion models, the model gradually reconstructs the original data from a random Gaussian noise.
\begin{equation}
p_{\theta}\left(x_{t-1} \mid x_{t}\right):=\mathcal{N}\left(x_{t-1} ; \tilde{\mu}_{t}, \tilde{\beta}_{t} \mathbf{I}\right),
\end{equation}
where variance \(\tilde{\beta}_{t}\) can be calculated from the forward process posteriors:
\begin{equation}
\tilde{\beta}_{t}:=\frac{1-\bar{\alpha}_{t-1}}{1-\bar{\alpha}_{t}} \beta_{t}
\end{equation}
According to \cite{DBLP:conf/icassp/HaiWYTDE24},
\begin{equation}
x_{0}:=\sqrt{\bar{\alpha}_{t}} x_{t}-\sqrt{1-\bar{\alpha}_{t}} v_{t}
\end{equation}
\begin{equation}
\tilde{\mu}_{t} =\frac{\sqrt{\bar{\alpha}_{t-1}} \beta_{t}}{1-\bar{\alpha}_{t}} x_{0}+\frac{\sqrt{\alpha_{t}}\left(1-\bar{\alpha}_{t-1}\right)}{1-\bar{\alpha}_{t}} x_{t}
\end{equation}

\begin{figure}[t]
  \centering
  \includegraphics[width=8.5cm]{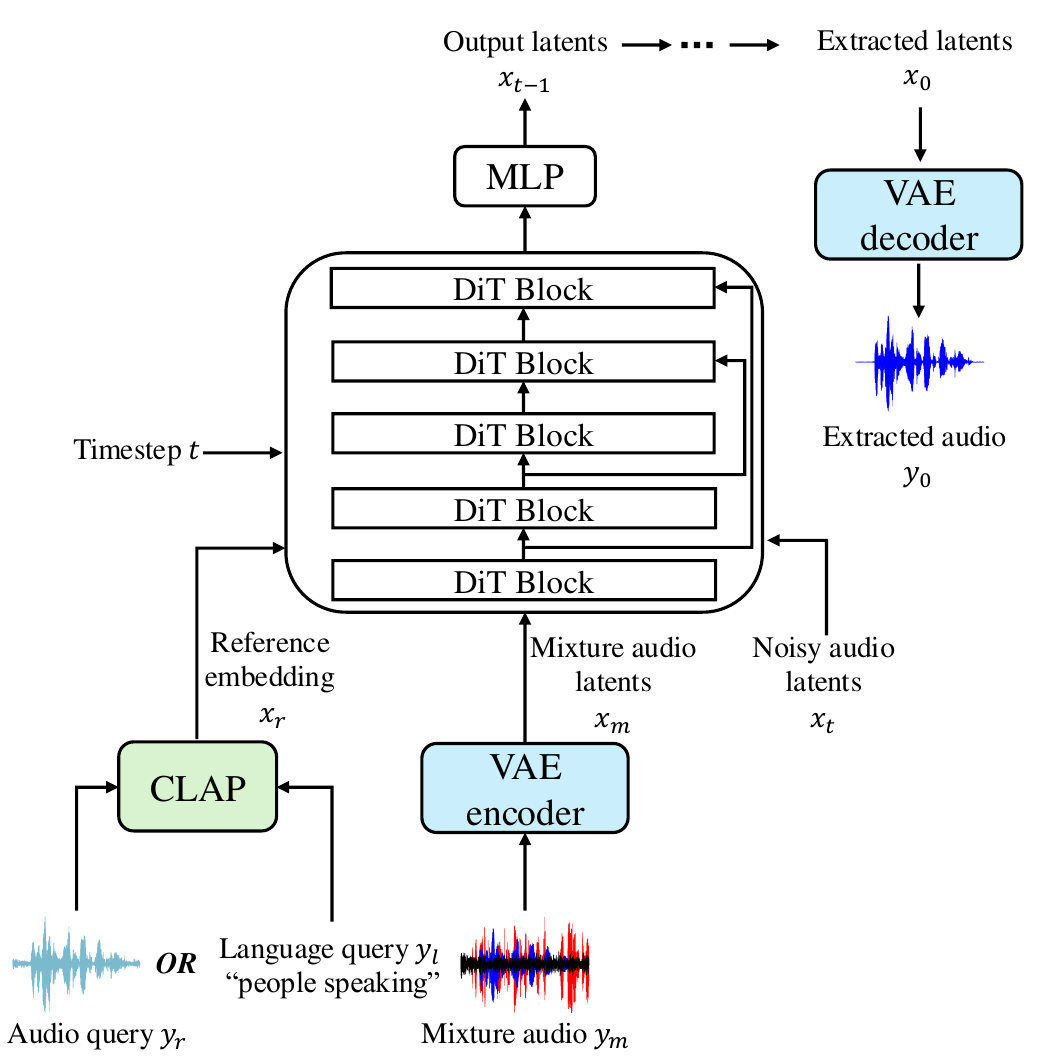}
  \caption{Diagram of SoloAudio model.
  }
  \label{fig:1}
\end{figure}

\subsection{SoloAudio}
As shown in Fig.~\ref{fig:1}, our proposed SoloAudio model consists of several key components: a VAE encoder, a VAE decoder, a CLAP model, and a DiT-like model \cite{DBLP:conf/iccv/PeeblesX23, DBLP:journals/corr/abs-2409-10819}. 

Given a 1-D mixture audio signal $y_m$, the VAE encoder is applied to extract audio latents $x_m \in \mathbb{R}^{N \times C}$, where $N$ represents the number of feature frames, and $C$ denotes the dimension of the latent channels. We leverage the VAE latent space for the diffusion process due to its superior reconstruction quality compared to the mel spectrogram space \cite{DBLP:conf/icml/EvansCTHP24}. The VAE model employs a fully-convolutional architecture, following the DAC encoder and decoder structure \cite{kumar2024high}, but with a VAE bottleneck rather than vector quantization.

The CLAP model, which bridges language and audio spaces and enables zero-shot predictions \cite{DBLP:conf/icassp/WuCZHBD23}, is used to extract the reference embedding $x_r$ from either an audio query $y_r$ or a language query $y_l$. For the noisy audio latents $x_t \in \mathbb{R}^{N \times C}$ at timestep $t$, we concatenate $x_t$ and $x_m$ on the channel dimension as the input to the DiT. 

\begin{figure}[t]
  \centering
  \includegraphics[width=4cm]{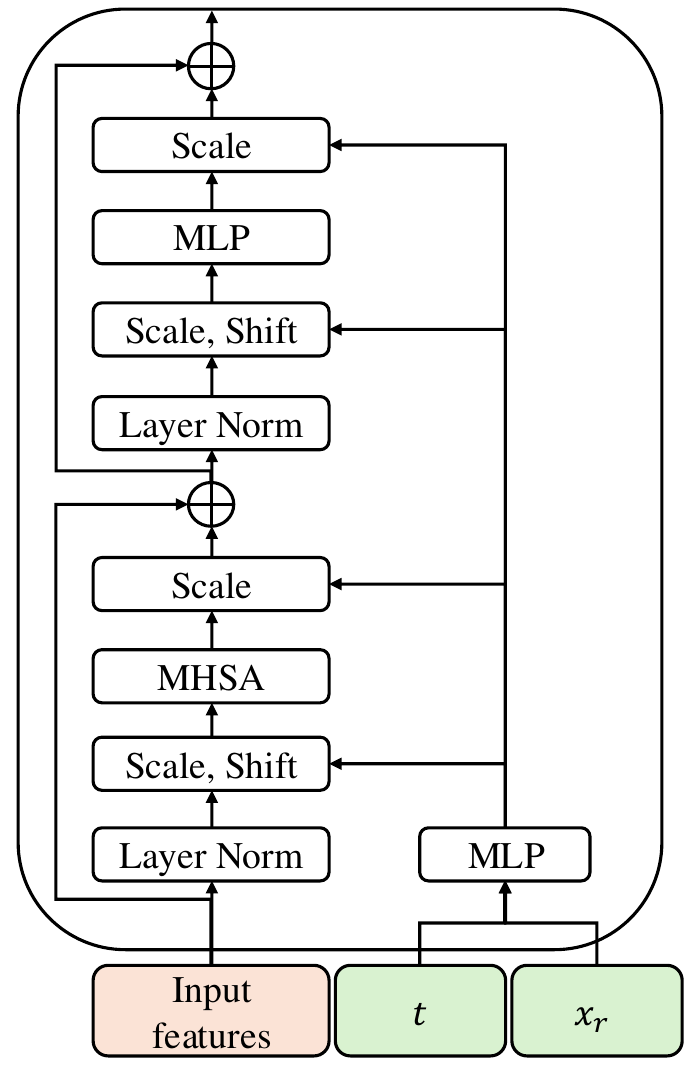}
  \caption{Diagram of the DiT block.
  }
  \label{fig:2}
\end{figure}

The DiT block, detailed in Fig.~\ref{fig:2}, includes an adaptive layer norm block, a multi-head self-attention (MHSA) block, and a multi-layer perceptron (MLP) block. The timestep $t$ and reference embedding $x_r$ serve as conditional information to regress the dimension-wise scale and shift parameters, which are incorporated into each block. 

The primary distinction between our network architecture and DiT lies in the use of long skip connections in SoloAudio, bridging shallow and deep DiT blocks as in \cite{DBLP:conf/cvpr/BaoNXCL0023}. These skip connections create shortcuts for low-level features, streamlining the training of the entire $v$-prediction network. Furthermore, we incorporate rotary positional embeddings (RoPE) \cite{DBLP:journals/ijon/SuALPBL24} for enhanced position encoding of audio latents.

\subsection{Inference}
During inference, we obtain the output latents $x_t \in \mathbb{R}^{N \times C}$ by feeding $x_t$, $y_m$, $y_r$ (or $y_l$) and $t$ into the DiT model. After $T$ denoising sampling steps, the clean target latents $x_0 \in \mathbb{R}^{N \times C}$ could be estimated. 

We apply classifier-free guidance (CFG) to steer the sampling process. This involves training the model in two modes: conditioned and unconditioned, enabling it to learn both how to generate general outputs and how to generate outputs that match specific conditioning inputs. The CFG technique adjusts the model's output $v_t$ during sampling, which can be expressed as:
\begin{equation}
    v^{\prime}_t = v^{uncond}_t +  \gamma (v^{cond}_t - v^{uncond}_t)
\end{equation}
where $\gamma$ represents the guidance scale, $v^{uncond}_t$ is the prediction of the unconditioned sampling and $v^{cond}_t$ is the prediction of the conditioned sampling.

\begin{table*}[t]
  \caption{Results on the FSD-mix dataset.}
  \label{tab:result1}
  \footnotesize
\renewcommand{\arraystretch}{1.1} 
  \centering
  \setlength{\tabcolsep}{10pt}
  \begin{tabular}{l|cccc|cccc}
    \hline
   \multirow{2}{*}{\textbf{Method}} & \multicolumn{4}{c|}{\textbf{Audio-oriented}} & \multicolumn{4}{c}{\textbf{Language-oriented}} \\
    \cline{2-9}
    &FD $\downarrow$& KL $\downarrow$&CLAP-audio $\uparrow$&ViSQOL $\uparrow$&FD $\downarrow$&KL $\downarrow$&CLAP-audio $\uparrow$&ViSQOL $\uparrow$\\
    \hline
    DPM-TSE &  29.262 & 1.661 & 0.623 & 2.180 &27.121&1.610&0.640&2.201\\
    \hline
    SoloAudio &\textbf{5.875} & \textbf{1.108} &\textbf{0.772} & \textbf{2.411}&\textbf{4.986}&\textbf{0.976}&\textbf{0.801}&\textbf{2.498}  \\
    \multicolumn{1}{r|}{\textit{w/o skip connection}}& 9.128 & 1.304  & 0.738 & 2.369 &8.481&1.170&0.763&2.457  \\
    \hline
  \end{tabular}
\end{table*}

\section{Experiments}
\subsection{DataSet}
\subsubsection{Synthetic data (FSD-Mix)}
Following \cite{DBLP:conf/interspeech/OchiaiDKIKA20, DBLP:conf/interspeech/WangYWYZ22}, we created datasets of simulated mixtures using the Freesound Dataset Kaggle 2018 corpus\footnote{https://www.kaggle.com/c/freesound-audio-tagging} (FSD) \cite{DBLP:conf/ismir/FonsecaPFFBFOPS17}. The audio clips in the FSD vary in length, from $0.3$ to $30$ seconds.
We generated $10$-second audio mixtures, each consisting of one target sound and $1$-$3$ interfering sounds, randomly selected from the FSD. The signal-to-noise ratio (SNR) of the interfering sounds is randomly set within a range of $-10$ to $10$ dB. These sounds were superimposed at random time points over a $10$-second background noise, sourced from the DCASE 2019 Challenge's acoustic scene classification task\footnote{https://dcase.community/challenge2019/task-acoustic-scene-classification} \cite{mesaros2018multi}. The SNR for the background noise was randomly set between $-5$ and $10$ dB.
All audio clips were resampled to $24$ kHz. Each training audio file was simulated for $3$ mixtures, resulting in $28,419$ samples for the training set, $160$ for the validation set, and $1,440$ for the test set.
The corpus contains $41$ sound event categories, ranging from human-produced sounds to musical instruments and object noises. 

\subsubsection{Synthetic data (TangoSyn-Mix)}
A recently released variant of Tango \cite{DBLP:conf/icassp/kong2024improving}, which has demonstrated state-of-the-art performance in text-to-audio generation, was used to synthesize data from text descriptions. Specifically, we used $300$ categories from VGG-Sound \cite{DBLP:conf/icassp/ChenXVZ20} and manually assessed the quality of the generated audio by listening to three samples from each category as Tango might fail to actually generate some sound categories. After initial filtering, $227$ categories were retained. For each category, we generated $24$ samples using different random seeds and text augmentations. The TangoSyn-Mix dataset was created following the same simulated process as the FSD-Mix dataset, resulting in a training set with a total of $95,340$ audio files. Compared to FSD-Mix, $22$ categories overlap with TangoSyn-Mix, while the remaining $19$ categories are excluded from TangoSyn-Mix and reserved for evaluating the few-shot and zero-shot capabilities of the models.

\subsubsection{Real evaluation data (AudioSet)}\label{real data}
The AudioSet evaluation set was used for real-world TSE evaluation \cite{DBLP:conf/icassp/GemmekeEFJLMPR17}. We selected audio from 41 FSD categories and randomly chose 5 samples per category. After listening to these samples, we manually selected 2 samples per category to ensure the presence of the category-related sound, resulting in a total of 82 selected audio samples.

We open-source the training and evaluation data used in our experiments.

\subsection{Experimental Setups}
We conducted experiments using a $24$kHz audio sample rate for both the waveform VAE and the SoloAudio model.
The waveform latent representation operates at 50Hz and contains 128 channels. The VAE was trained on AudioSet to handle a wide range of general audio classes.
SoloAudio’s DiT follows DiT-B\footnote{https://github.com/facebookresearch/DiT/blob/main/models.py}, which is composed of $12$ DiT blocks, each with $768$ channels and $12$ attention heads.

The CLAP embedding has a dimension of 512. We augment the text using the following formats: ``[CLS]", ``\textit{An audio clip of }[CLS]", or ``\textit{The sound of }[CLS]", where [CLS] is the target sound category.
Our model was trained using the AdamW optimizer with a learning rate of $0.0001$, weight decay of $0.0001$, a batch size of $128$, and for $100$ epochs. The diffusion and inference steps for SoloAudio are set to $1000$ and $50$, respectively, with the variance $\beta$ ranging from $0.00085$ to $0.012$. 
The model was trained on one NVIDIA A100-80GB GPU for two days.
We allocated $10\%$ of the data for unconditioned training and $90\%$ for conditioned training. During sampling, the default guidance scale $\gamma$ is set to $2.5$ for audio-oriented TSE and $3.0$ for language-oriented TSE based on our ablation studies.
For the few-shot experiments, we fine-tuned the model using the AdamW optimizer with a learning rate of $0.00001$, a weight decay of $0.0001$, and a batch size of $32$ over $20$ epochs.

\subsection{Baselines}
We compare SoloAudio with three modern TSE models: WaveFormer\footnote{https://github.com/vb000/Waveformer} \cite{DBLP:conf/icassp/VeluriCICYG23}, AudioSep\footnote{https://github.com/Audio-AGI/AudioSep} \cite{DBLP:journals/corr/abs-2308-05037}, and DPM-TSE\footnote{https://github.com/haidog-yaqub/DPMTSE} \cite{DBLP:conf/icassp/HaiWYTDE24}. WaveFormer operates in the waveform domain, AudioSep works on the STFT representation, and DPM-TSE uses the mel-spectrogram. Both WaveFormer and AudioSep support text-oriented TSE, and due to limited computational resources, we directly use their official checkpoints for the real-world TSE evaluation. WaveFormer was trained on the FSD mixture dataset, while AudioSep was trained on large-scale audio mixtures. DPM-TSE is originally designed to use one-hot labels; we retrained it by substituting the one-hot embeddings with CLAP embeddings for both audio-oriented and language-oriented TSE.


\subsection{Metrics}

Following \cite{DBLP:conf/icassp/HaiWYTDE24}, we introduce perceptual evaluations and subjective assessment to evaluate TSE models.

\subsubsection{Objective metrics}
We use five automatic evaluation functions:
(i) \textbf{ViSQOL} \cite{chinen2020visqol} is an algorithm to assess the quality of audio signals by approximating human perceptual responses based on five-scaled mean opinion scores.
(ii) \textbf{Frechet Distance (FD)} \cite{DBLP:conf/icml/LiuCYMLM0P23} in audio indicates the similarity between
generated samples and target samples. 
(iii) \textbf{Kullback–Leibler (KL) divergence} is measured
at a paired sample level and averaged as the final result.
FD and KL are built upon a state-of-the-art audio classifier PANNs \cite{DBLP:journals/taslp/KongCIWWP20}.
(iv) \textbf{CLAP-audio} is calculated using CLAP features between generated samples and target samples.
(v) \textbf{CLAP-text} is calculated using CLAP features between generated samples and target text.

\subsubsection{Subjective metrics}
Following \cite{DBLP:conf/icassp/HaiWYTDE24}, we recruited $12$ participants with recording or music production experiences to evaluate the listening perceptual quality of audios predicted by different TSE models.
We evaluated the performance of language-oriented TSE in real-world application scenarios using the real evaluation data described in Section \ref{real data}. Each subject was asked to evaluate 41 audio pairs for each model. Each audio pair included the original mixture, a description of the target sound, and the model’s prediction for the extracted sound.
For each audio pair, subjects were asked to respond to two questions:

\textbf{(i) Extraction: Does the generated audio contain the target sound as described in the text?}  
Ratings ranged from $1$ to $5$, where $1$ indicated that the target sound could not be heard at all in the generated audio, and $5$ indicated that the generated audio fully captured the target sound from the mixture as described.

\textbf{(ii) Purity: Does the generated audio only contain the sound corresponding to the text description?}  
Ratings ranged from $1$ to $5$, where $1$ indicated that the generated audio contained many unrelated sounds, and $5$ indicated that it contained only the target sound with no detectable unrelated sounds.

\begin{figure*}[t]
    \centering
    \subfigure[FD $\downarrow$]{\includegraphics[width=0.24\textwidth]{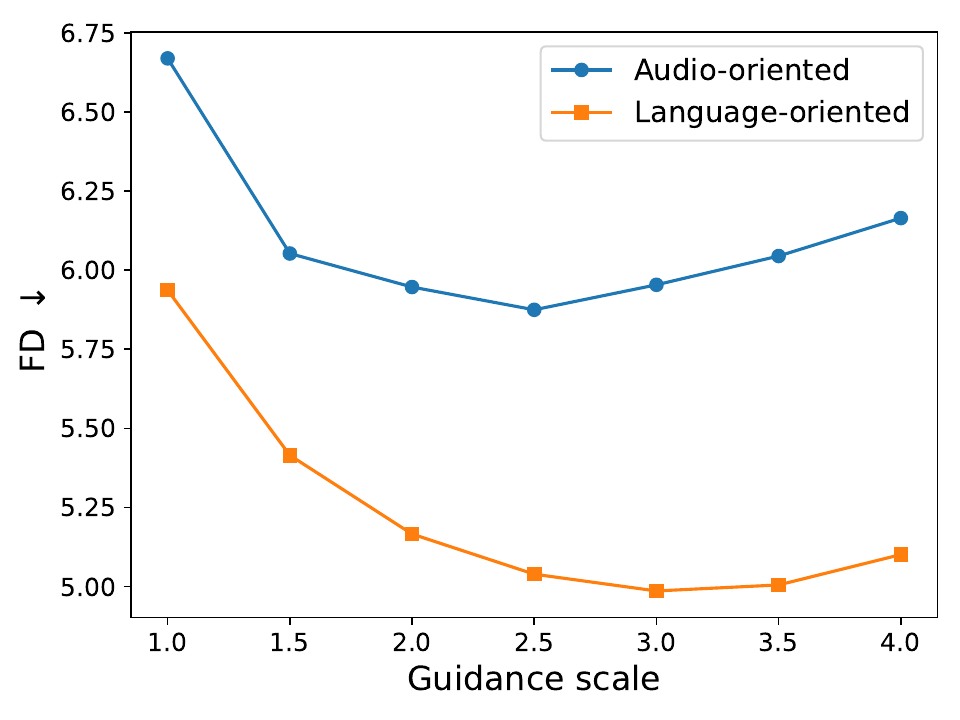}}
    \subfigure[KL $\downarrow$]{\includegraphics[width=0.24\textwidth]{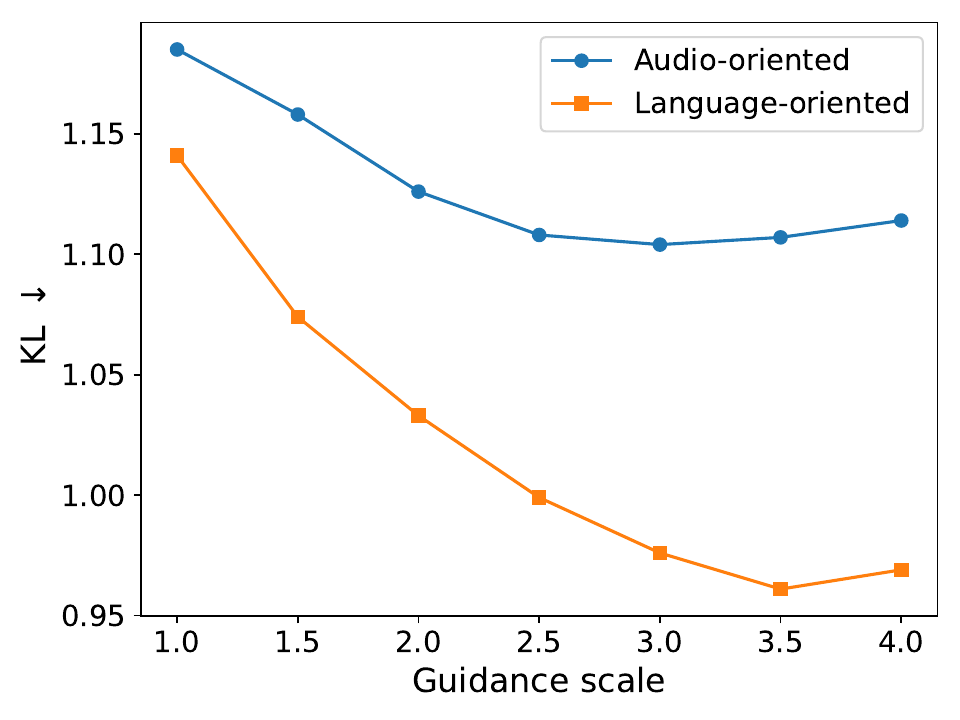}}
    \subfigure[CLAP-audio $\uparrow$]{\includegraphics[width=0.24\textwidth]{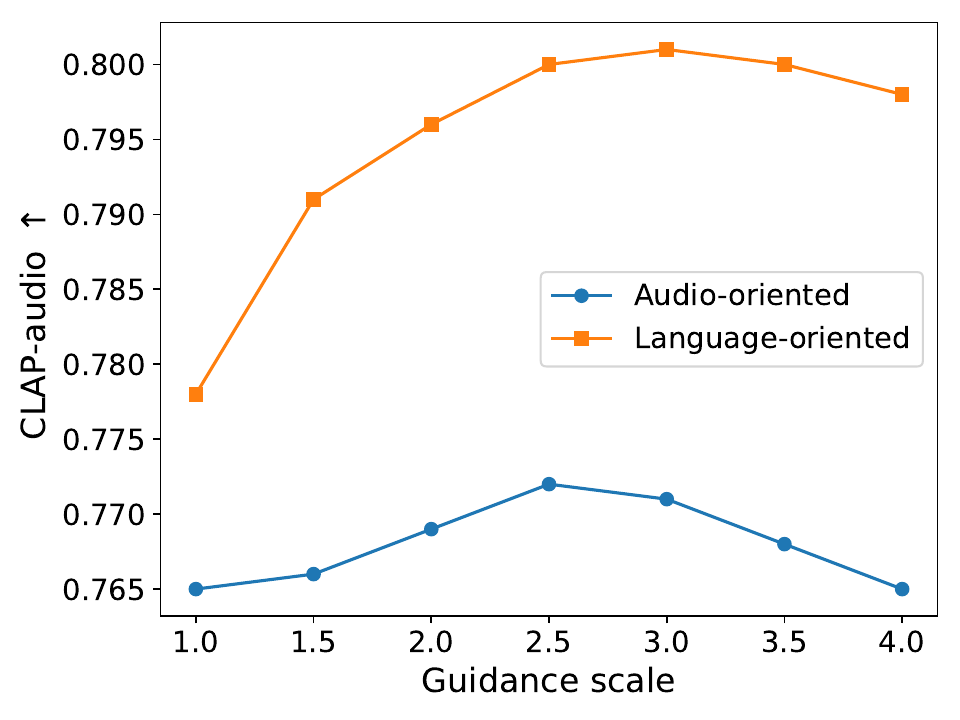}}
    \subfigure[ViSQOL $\uparrow$]{\includegraphics[width=0.24\textwidth]{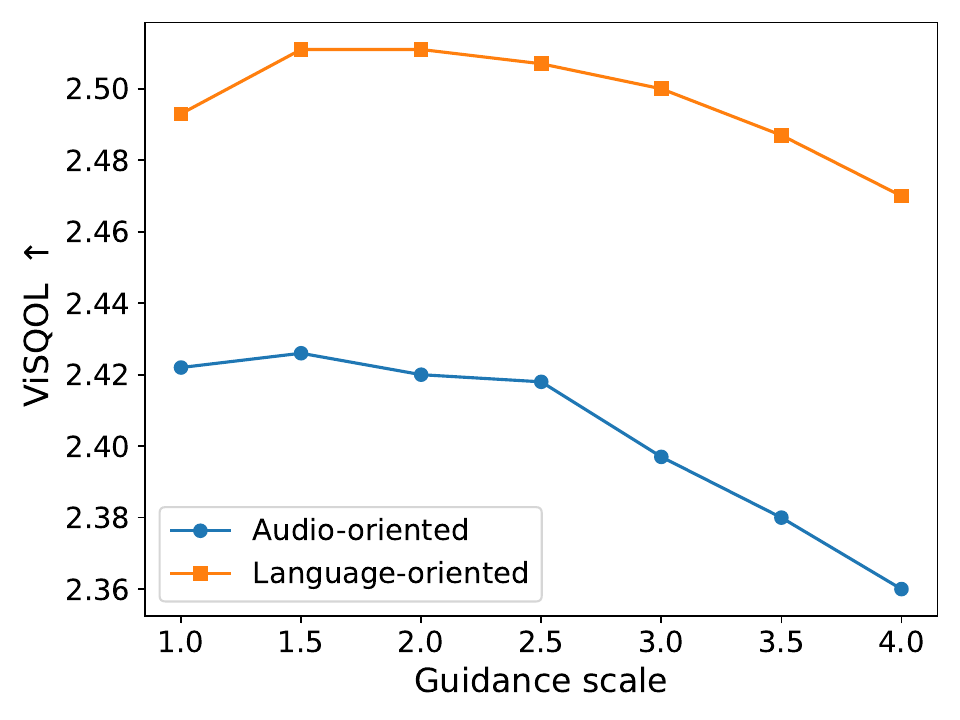}}
    \caption{Influence of the guidance scale}
    \label{fig:3}
\end{figure*}


\begin{table*}[t]
  \caption{Results on the FSD-mix dataset. We test both 22 seen labels (S) and 19 unseen labels (UNS) from the SynVGG-mix training data.}
  \label{tab:result2}
  \renewcommand{\arraystretch}{1.25} 
  \footnotesize
  \centering
  \scriptsize
  \begin{tabular}{c|cccccccc|cccccccc}
    \hline
  \multirow{4}{*}{\textbf{Method}} & \multicolumn{8}{c|}{\multirow{2}{*}{\textbf{Audio-oriented}}} & \multicolumn{8}{c}{\multirow{2}{*}{\textbf{Language-oriented}}} \\
    & & & & & & & & & & & &  & & & &\\
    \cline{2-17}
    &\multicolumn{2}{c}{FD $\downarrow$}& \multicolumn{2}{c}{KL $\downarrow$}&\multicolumn{2}{c}{CLAP-audio $\uparrow$}&\multicolumn{2}{c|}{ViSQOL $\uparrow$}&\multicolumn{2}{c}{FD $\downarrow$}&\multicolumn{2}{c}{KL $\downarrow$}&\multicolumn{2}{c}{CLAP-audio $\uparrow$}&\multicolumn{2}{c}{ViSQOL $\uparrow$}\\
    &S&UNS&S&UNS&S&UNS&S&UNS&S&UNS&S&UNS&S&UNS&S&UNS\\
    \hline
    FSD&7.015&8.171&1.015&1.198&0.787&0.757&2.453&2.370&6.406&7.192&0.882&1.069&0.812&0.790&2.523&2.473\\
     FSD+TangoSyn &5.317& 6.158&0.879& 1.132&0.806& 0.771&2.498& 2.398& 4.779&5.048 & 0.676&0.818 & 0.844&0.827 & 2.630&2.583\\

    \hline
    zero-shot & 22.673 &20.303&1.983&2.220&0.594&0.576&2.003&2.050&39.884&38.009&1.685&2.333&0.641&0.570&2.146&2.014 \\
    one-shot & 14.144&13.121&1.610&1.890&0.644&0.619&2.161&2.124&9.461&12.368&1.338&1.975&0.718&0.623&2.319&2.114 \\
    10-shot & 8.361 &9.894&1.359&1.671&0.734&0.677&2.383&2.230&7.852&10.388&1.129&1.666&0.758&0.671&2.437&2.220\\
    \hline
  \end{tabular}
\end{table*}

\begin{table}[t]
  \caption{Results on the real AudioSet dataset. We report Extraction and Purity results with their $95\%$ confidence intervals.}
  \label{tab:result3}
  \footnotesize
  \renewcommand{\arraystretch}{1.25} 
  \centering
  \begin{tabular}{c|ccc}
    \hline
    \textbf{Method} & \textbf{CLAP-text} $\uparrow$&\textbf{Extraction $\uparrow$} &\textbf{Purity $\uparrow$} \\
    \hline
    AudioSep & $0.168$&$\boldsymbol{4.431}\pm0.089$&$2.487\pm0.142$\\
    WaveFormer & $0.097$&$3.492\pm0.109$&$3.266 \pm0.132$\\
    DPM-TSE &  $0.096$&$2.437\pm0.123$&$3.939 \pm0.130$\\
    \hline
    SoloAudio & $\boldsymbol{0.213}$&$3.923\pm0.113$&$\boldsymbol{4.263} \pm0.109$ \\
    \hline
  \end{tabular}
\end{table}

\section{Results}
\subsection{Comparison with DPM-TSE}
We compare SoloAudio with DPM-TSE using in-domain data, training and testing both models on the FSD-Mix dataset under identical conditions. As shown in Table~\ref{tab:result1}, SoloAudio significantly outperforms DPM-TSE across all metrics. Both audio-oriented and language-oriented TSE highlight the effectiveness of SoloAudio.
Besides, we found that the language-oriented TSE performs better thant the audio-oriented TSE.

\subsection{Ablation Studies}
Table~\ref{tab:result1} shows the impact of adding skip connections to the DiT model, resulting in a clear performance improvement. 
In addition, we examine the impact of the CFG guidance scale on model performance. As shown in Fig.~\ref{fig:3}, as the guidance scale increases, performance initially improves but then declines. We select optimal values of $2.5$ for the audio-oriented TSE and $3.0$ for the language-oriented TSE.

\subsection{Influence of Synthetic Data}
We compare the results of SoloAudio on FSD-mix data with and without synthetic data.
The FSD data contains 22 labels present in the TangoSyn data, leaving 19 labels unseen.
Table~\ref{tab:result2} highlights the impact of synthetic data, showing that using TangoSyn clearly improves TSE performance on both seen and unseen data.

\subsection{Zero-shot and Few-shot TSE}
To further evaluate the few-shot and zero-shot capabilities of the models, we utilized the SoloAudio model trained exclusively on TangoSyn data. For the zero-shot setting, we directly tested the model on the out-of-domain FSD-Mix test set, which contains unseen labels. In the few-shot setting, we fine-tuned the model using either $1$ or $10$ samples per category from the FSD-Mix training set and evaluated its performance on the FSD-Mix test set. Table~\ref{tab:result2} presents these results. Overall, SoloAudio demonstrates remarkable zero-shot capability on out-of-domain data with unseen labels. Moreover, fine-tuning with a small number of samples ($1$ or $10$) leads to a significant performance improvement across all metrics.

\subsection{Performance on Real Data}
Furthermore, we performed both objective and subjective evaluations on real data to compare SoloAudio with three state-of-the-art TSE models. Table~\ref{tab:result3} summarizes the performance of the models, with our proposed SoloAudio achieving the highest CLAP-text score, demonstrating strong alignment with the target sound prompt. In the listening test, SoloAudio records the highest Purity score and a strong Extraction score, highlighting its clear advantage in isolating and recovering target sounds with minimal interference. Although AudioSep achieves the highest Extraction score, its low Purity score indicates difficulties in removing unrelated noise. This issue could arise from training the model on multi-label audio samples, which may hinder its ability to accurately extract individual sounds.

\section{Conclusions}
In this paper, we propose a generative method for TSE, built on a latent diffusion model with a skip-connected Transformer. We also explore the use of synthetic data generated by T2A, demonstrating its strong potential for training TSE models. In future work, we aim to (1) improve the sampling speed of SoloAudio, (2) investigate more effective T2A tools and audio-text alignment methods, (3) scale up training with larger datasets, and (4) explore the use of alternative target references, such as images and videos.

{

\bibliographystyle{IEEEtran}
\bibliography{IEEEexample}
}

\end{document}